\documentclass[artice,aps,pra,showpacs,twocolumn,superscriptaddress]{revtex4}
\usepackage{graphicx,color}
\usepackage{amsmath,amsthm,amsfonts,amssymb,bm}
\usepackage{times}
\usepackage{epsf}
\usepackage[colorlinks={true}]{hyperref}
\usepackage[T1]{fontenc}
\usepackage[utf8]{inputenc}
\hypersetup{citecolor={blue}, filecolor={blue}, linkcolor={blue}, urlcolor={blue}}

\newcommand{\commentold}[1]{}
\DeclareMathSymbol{:}{\mathpunct}{operators}{"3A}
\usepackage[utf8]{inputenc}
\usepackage[english]{babel}
\usepackage{amsthm}
\theoremstyle{definition}

\begin{document}

\title{Holevo bound of entropic uncertainty relation under Unruh channel in the context of
open quantum systems}
\author{S. Haseli}
\email{soroush.haseli@uut.ac.ir}
\affiliation{Department of Physics, Urmia University of Technology, Urmia, Iran
}
\author{F. Ahmadi}
\affiliation{Department of Engineering Science and Physics, Buein Zahra Technical University
}


\date{\today}

\begin{abstract}
The uncertainty principle is the most important feature of quantum mechanics, which can be called the heart of quantum mechanics. This principle sets a lower bound on the uncertainties of two incompatible measurement. In quantum information theory, this principle is expressed in terms of entropic measures. Entropic uncertainty bound can be altered by considering a particle as a quantum memory. In this work we investigate the entropic uncertainty relation under the relativistic motion. In relativistic  uncertainty game Alice and Bob agree on two observables, $\hat{Q}$ and $\hat{R}$, Bob prepares a particle constructed from the free fermionic mode in the quantum state and sends it to Alice, after sending, Bob
begins to move with an acceleration $a$, then Alice does a measurement on her particle $A$ and announces her choice to Bob, whose task is then to minimize the uncertainty about the measurement outcomes. we will have an inevitable increase  in the uncertainty of the Alic's measurement outcome due to information loss which was stored initially in B. In this work we look at the Unruh effect as a quantum noise and we will characterize it as a quantum channel. 
\end{abstract}
\maketitle
\maketitle

\section{Introduction}
The distinction between quantum theory and classical theory is formulated by Heisenberg’s uncertainty relation \cite{Heisenberg}. It provides a certain limitation on our ability  to predict the precise outcomes of two incompatible measurements on a quantum system. In the original, for arbitrary pairs of noncommuting observables $\hat{Q}$ and $\hat{R}$ and quantum state $\vert \psi \rangle$, the uncertainty relation is introduced in terms of the standard deviation \cite{Robertson,Schrodinger}
\begin{equation}\label{standard}
\Delta Q \Delta R \geq \frac{1}{2} \vert \langle \psi \vert \left[ Q,R \right] \vert \psi \rangle \vert,
\end{equation} 
where $\Delta Q = \sqrt{ \langle \psi \vert Q^{2} \vert \psi \rangle - \langle \psi \vert Q \vert \psi \rangle^{2}}$ and $\Delta R = \sqrt{ \langle \psi \vert R^{2} \vert \psi \rangle - \langle \psi \vert R \vert \psi \rangle^{2}}$ are the standard deviations and $\left[ Q,R \right] = Q~R-R~Q$. This uncertainty relation is state dependent and trivial  when the  state is one of the eigenstates of the observables \cite{Prevedel}.  It is more natural to quantify uncertainty in terms of entropy rather than the standard deviation. Bialynicki-Birula and Mycielski \cite{Mycielski}
derived entropic uncertainty relation (EUR) for position and
momentum and Deutsch \cite{Deutsch} later proved a relation that
holds for any pair of observables. Subsequently, Kraus \cite{Kraus} 
conjectured an improvement of Deutsch's result which was later proven by Maassen and Uffink \cite{Maassen}. It states that, for two incompatible observable $\hat{Q}$ and $\hat{R}$ with two sets of eigenbases $ \lbrace \vert q_{i} \rangle \rbrace$ and $\lbrace \vert r_{j}\rangle \rbrace$ we have
\begin{equation}\label{entropic}
H(Q)+H(R) \geq \log_{2}\frac{1}{c},
\end{equation} 
where $H(X)=-\sum_{x} p_{x} \log_{2} p_{x}$ is the Shannon entropy of the measured observable $X \in \lbrace Q, R\rbrace$, $p_x$  is the probability of
the outcome $x$, and $c=\max_{\lbrace i,j\rbrace} \vert \langle q_i \vert r_j \rangle \vert^{2}$ quantifies the ‘complementarity’ between the observables. In general, For measured particle with density matrix $\rho$, Eq. \ref{entropic} can be expressed as\cite{Frank,Berta}
\begin{equation}
H(Q)+H(R) \geq \log_2 \frac{1}{c}+S(\rho)
\end{equation} 
where $S(\rho)=-tr(\rho \log_2 \rho)$ is the von Neumann entropy of $\rho$. Another scenario that can be considered is the presence of another particle as the quantum memory. This scenario is described by Berta et al. \cite{Berta}, they consider a quantum memory $B$ which has correlation with measured particle $A$. They showed that the uncertainty of Bob, who has access to the quantum memory, about the result of measurements $\hat{Q}$ and $\hat{R}$ on the Alice’s particle, $A$, Satisfies the following EUR
\begin{equation}\label{berta}
S(Q \vert B)+ S(R \vert B) \geq \log_2 \frac{1}{c} + S(A \vert B),
\end{equation}
where $S(O \vert B)=S(\rho^{OB})-S(\rho^{B})$, $O \in \lbrace Q,R \rbrace$ the conditional von Neumann entropies of the post measurement states
\begin{equation}
\rho^{OB}=\sum_{i} (\vert o_i \rangle \langle o_i \vert \otimes \mathbf{I})\rho^{AB} (\vert o_i \rangle \langle o_i \vert\otimes \mathbf{I}),
\end{equation}
where $\lbrace \vert o_i \rangle\rbrace$’s are the eigenstates of the observable $\hat{O}$, and $\mathbf{I}$ is the identity operator. This (EUR) has many applications within entanglement detection \cite{Huang,Prevedel,Chuan-Feng} and quantum cryptography \cite{Tomamichel,Ng}. So far, many efforts have been made to improve Berta's (EUR) \cite{Pati,Pramanik,Coles,Liu,Zhang,Pramanik1,Adabi,Adabi1,Dolatkhah,Jin-Long}. In Ref. \cite{Adabi} Adabi et al. obtained another lower bound for EUR in the presence of quantum memory. They consider a bipartite state $\rho_{AB}$ sharing between Alice and Bob. Alice performs $\hat{Q}$ or $\hat{R}$ measurement and announce her choice to Bob. Bob’s uncertainty about both $\hat{Q}$ and $\hat{R}$ measurement outcomes is \cite{Adabi}
\begin{eqnarray}
S(Q\vert B)+S(R\vert B)&=&\\ \nonumber
&=&H(Q)-I(Q;B)+H(R)-I(R;B)\\ \nonumber
&\geq&\log_2 \frac{1}{c}+S(A)-[I(Q;B)+I(R;B)]\\ \nonumber
&=&\log_2 \frac{1}{c} + S(A\vert B)+ \\ \nonumber
&+& \lbrace I(A;B)- [I(Q;B)+I(R;B)]\rbrace,
\end{eqnarray}
where in the second line, we use Eq. \ref{entropic} and in the last
line we apply $S(A)=S(A \vert B)+ I(A;B)$, thus the EUR is obtained as
\begin{equation}\label{concen}
S(Q\vert B)+S(R\vert B) \geq \log_2 \frac{1}{c} + S(A \vert B)+ \max \lbrace 0, \delta\rbrace,
\end{equation}
where 
\begin{equation}
\delta = I(A;B)-(I(Q;B)+I(R;B)),
\end{equation}
thus the uncertainties $S(Q \vert B)$ and $S(R \vert B)$ are lower bounded by an additional term compared to Berta's uncertainty relation. Note that, when Alice measures observable $P$, the $i$-th outcome with probability $p_i= tr_{AB}(\Pi_i^{A} \rho^{AB} \Pi_i^{A}) $ is obtained and Bob state  is left in the corresponding state $\rho_i^{B}=\frac{tr_{A}(\Pi_i^{A} \rho^{AB} \Pi_i^{A})}{p_i}$, thus 
\begin{equation}
I(P;B)=S(\rho^{b})-\sum_{i}p_{i}S(\rho_{i}^{B})
\end{equation}
is the Holevo quantity and it is equal to the upper bound of the Bob accessible information about the Alice measurement outcomes. From here on, we call this (EUR) as Holevo (EUR). 

Decoherence is an inevitable phenomenon related to open quantum systems. It occurs
due to the interaction of the system with the environment. Decoherence leads to decay of quantum correlations and quantum coherence which are important resources for quantum information processing. It is known that  a quantum system observed in an accelerated reference frame  is similar to a quantum system which is influenced by quantum noise \cite{Omkar}. 

Here we study the effects of the  Unruh-effect \cite{Unruh} on the (EUR) for a Dirac field mode. We consider the bipartite scheme in which the quantum information
is shared between an inertial observer (Alice) and an accelerated
observer (Bob) in the case of Dirac field. In our uncertainty game, Alice and Bob firstly agree on two observables, $\hat{Q}$ and $\hat{R}$ then Bob prepares
a particle in a quantum state and sends it to Alice. After sending, Bob
begins to move with an acceleration $a$ and finally Alice does a measurement on her particle in one of the two agreed bases and tells her choice to Bob, whose task is then to minimize the uncertainty about the measurement outcomes. In  this situation, the  Von-Neumann entropy in (EUR) is observer dependent\cite{Marolf}. One can show that the quantum information stored initially in $B$ would be degraded in an accelerated reference frame,  this leads to an  increase of the entropic uncertainty lower bound (EULB) of the Alice's measurements  outcome. The remainder of this paper is organized as follows.  In Sec. \ref{UNRUH EFFECT} we introduce the notion of the Unruh effect for a
two-mode fermionic system. In Sec. \ref{GEO} we specified the Unruh effect as a quantum noise channel. We give some examples in Sec. \ref{exam}. The manuscript closes with conclusion in Sec. \ref{con}.  
\section{UNRUH EFFECT}\label{UNRUH EFFECT}
The Unruh effect states that the Minkowski vacuum as seen by an observer accelerating with constant  acceleration $a$ will arise as a warm gas emitting black-body radiation at the Unruh temperature \cite{Davies,Crispino}
\begin{equation}
\tau = \frac{\hbar a}{2 \pi K_B c},
\end{equation} 
where $c$ is the speed of light in vacuum, and $K_B$ is Boltzmann's constant. The Unruh effect leads to a decoherence-like effect, thus one can consider the Unruh effect as quantum noise. It is therefore considered as a quantum channel called "Unruh Channel". It affects on the quantum information shared between two observer Alice and Bob, in the case of bosonic or Dirac field mode\cite{Alsing,Alsing1}. 

Things that have been done so far in the field of Unruh effect is concentrated on either the bosonic \cite{Fuentes} or fermionic \cite{Alsing} channels, depending
on whether one is working with a scalar or a Dirac field, respectively. Due
to the finite occupation of the fermionic states, one can obtain finite-dimensional density matrices that lead to closed-form expressions for quantum information quantity that are more easily interpreted than their infinite-dimensional bosonic scalar field. This motivates us to consider the effect of other aspects of quantum information on the fermionic Unruh channel.
The Unruh effect is defined by investigating the Minkowski space time in terms of
Rindler coordinates\cite{Omkar,Unruh,Rindler}. The Rindler transformation divides spacetime into two disconnected regions, such that, a uniformily accelerated observer in one region is separated from the other region\cite{Alsing}. Since the field modes restricted in different region are not connected, the quantum information of accelerated observer degrades and leads to thermal bath.  In this work we consider the  fermionic field with few degree, as is in Ref. \cite{Feng}.

Consider two observers, Alice (A) and Bob (B) sharing a maximally entangled state of two Dirac field modes at Minkowski spacetime. Thus the quantum field is in a state
\begin{equation}\label{mink}
\frac{1}{\sqrt{2}}(\vert 0_s\rangle^{\mathcal{M}}  \vert 0_k \rangle^{\mathcal{M}}  + \vert 1_s\rangle^{\mathcal{M}}  \vert 1_k \rangle^{\mathcal{M}}  ),
\end{equation}
the states $\vert 0_j \rangle^{\mathcal{M}}$ and $\vert 1_j \rangle^{\mathcal{M}}$ are the vacuum and single
particle excitation states of the mode j in Minkowski space. We assume Alice's state constructed by the field mode s and she has a detector which only
detects mode s, while Bob’s states can be created from k field modes and has a detector sensitive only
to mode k. If Bob  uniformly undergoes acceleration $a$, the states corresponding to field mode k must be defined in Rindler bases in order to describe what Bob sees. By applying the presented formalism the Minkowski vacuum state transforms to the Unruh mode(a two-mode squeezed state), however the excited state is a product state, these state in terms of modes in different Rindler region $I$ and $II$ can be expressed as\cite{Alsing}, 
\begin{eqnarray}
\vert 0_s\rangle^{\mathcal{M}}&=&\cos r \vert 0_k \rangle_{I} \vert 0_k \rangle_{II}+\cos r \vert 1_k \rangle_{I} \vert 1_k \rangle_{II}\\ \nonumber
\vert 1_s\rangle^{\mathcal{M}}&=&\vert 1_k \rangle_{I} \vert 0_k \rangle_{II},
\end{eqnarray}
where $\cos r = \frac{1}{\sqrt{1+e^{\frac{-2 \pi \omega c}{a}}}}$, $\omega$ is Dirac particle frequency with acceleration $a$ and $c$ is the speed of light in vacuum. Note that $a \in \left[0,\infty \right] $ and thus, $\cos r \in \left[ \frac{1}{\sqrt{2}},1\right]$. By using this formalism, the state in Eq. \ref{mink} transform to
\begin{eqnarray}
\vert \psi \rangle &=&\frac{1}{\sqrt{2}}(\vert 0_s\rangle^{\mathcal{M}}(\cos r \vert 0_k \rangle_{I} \vert 0_k \rangle_{II}+ \\ \nonumber 
&+&\cos r \vert 1_k \rangle_{I} \vert 1_k \rangle_{II})
+\vert 1_s\rangle^{\mathcal{M}}\vert 1_k \rangle_{I} \vert 0_k \rangle_{II} ),
\end{eqnarray}

Due to the fact that the region $I$ and $II$ are  disconnected regions, in Rindler's spacetime, the region $II$ can be traced out to obtain the following density matrix
\begin{eqnarray}
\rho^{AI}&=&\frac{1}{2}[\cos^{2}r\vert 0_{s}^{\mathcal{M}},0_{k}^{I}\rangle \langle 0_{s}^{\mathcal{M}},0_{k}^{I}\vert  + \\ \nonumber
&+&  \cos r(\vert 0_{s}^{\mathcal{M}},0_{k}^{I}\rangle \langle 1_{s}^{\mathcal{M}},1_{k}^{I}\vert + \vert 1_{s}^{\mathcal{M}},1_{k}^{I}\rangle \langle 0_{s}^{\mathcal{M}},0_{k}^{I}\vert)+\\ \nonumber
&+& \sin^{2}r \vert 0_{s}^{\mathcal{M}},1_{k}^{I}\rangle \langle 0_{s}^{\mathcal{M}},1_{k}^{I}\vert +\vert 1_{s}^{\mathcal{M}},1_{k}^{I}\rangle \langle 1_{s}^{\mathcal{M}},1_{k}^{I}\vert].
\end{eqnarray}
Here, the transformation of Bob's state into a mixed state is called a the Unruh channel
for a fermionic qubit. We discuss about this context in the following. 
\section{GEOMETRIC CHARACTERIZATION OF THE UNRUH CHANNEL}\label{GEO}

In this section  the unruh channel is characterized from the dynamical maps view point. Here we use the Choi-Jamiolkowski isomorphism for unruh channel. Choi matrix is obtained by applying the Unruh channel on one half of a maximally entangled two-qubit state\cite{Choi,Jamiolkowski}
\begin{equation}
\rho_U=
  \left[ {\begin{array}{cccc}
   \cos^2 r & 0 & 0 & \cos r \\
   0 & \sin^2 r & 0 & 0 \\
   0 & 0 & 0 & 0 \\
   \cos r & 0 & 0 & 1 \\
  \end{array} } \right],
\end{equation}
$\rho_U$ is the Choi matrix corresponding to Unruh channel $\varepsilon_U$. Finally, one can obtain the Kraus operators by diagonalizing the Choi matrix and the Unruh channel is characterized completely\cite{Usha}
\begin{equation}
 K_1=
  \left[ {\begin{array}{cc}
   \cos r & 0 \\
   0 & 1 \\
  \end{array} } \right],  \quad K_2=
  \left[ {\begin{array}{cc}
   0 & 0 \\
   \sin r & 0 \\
  \end{array} } \right],
\end{equation}
so that
\begin{equation}
\varepsilon_U(\rho)=\sum_{j=1,2}K_j \rho K_j^{\dag}
\end{equation}
with the completeness condition 
\begin{equation}
\sum_{j=1,2}K_j^{\dag} K_j = \mathcal{I} 
\end{equation}
As is clear from this Kraus representation the Unruh channel is similar to the amplitude damping channel ,which shows the effect of a zero temperature thermal bath \cite{Nielsen,Banerjee}. However Unruh effect corresponds to a finite temperature and it is expected to correspond to
the generalized amplitude damping  channels, which is a finite temperature channels.
\begin{figure}[t]
\includegraphics[scale=.53]{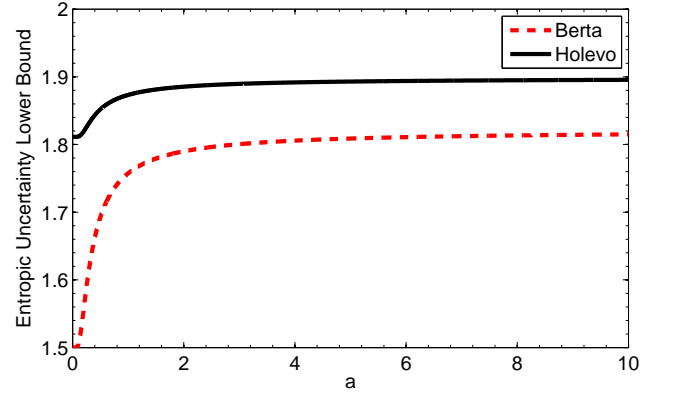}
\caption{(Color online )Entropic uncertainty lower bounds in terms of the  Bob's acceleration when the initial state $\rho_{AB}=p \vert \psi^- \rangle\langle \psi^- \vert + \frac{1-p}{2}(\vert \psi^+ \rangle\langle \psi^+ \vert + \vert \phi^+ \rangle\langle \phi^+ \vert)$  is shared between Alice and Bob. We consider the two complementary observable $\lbrace \sigma_x, \sigma_y \rbrace$, $p=\frac{1}{2}$ and $\omega=0.1$ }
\label{Fig1}
\end{figure} 
\begin{figure}[t]
\includegraphics[scale=.53]{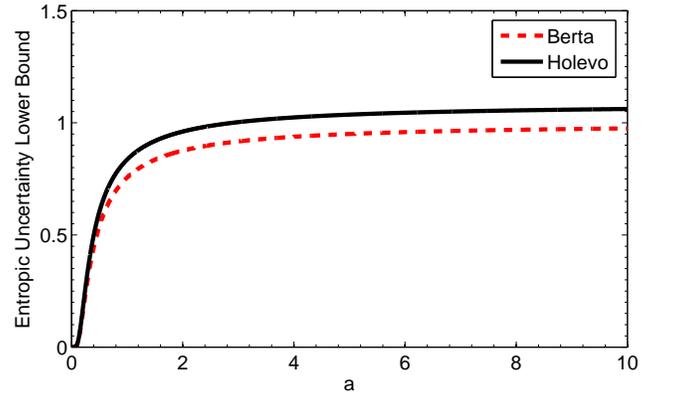}
\caption{(Color online )Entropic uncertainty lower bounds in terms of the  Bob's acceleration when the initial state $\rho^{AB}=p \vert \psi^{+}\rangle \langle \psi^{+} \vert + (1-p)\vert 11 \rangle \langle 11 \vert \vert$  is shared between Alice and Bob. We consider the two complementary observable $\lbrace \sigma_x, \sigma_y \rbrace$, $p=1$ and $\omega=0.1$.}
\label{Fig2}
\end{figure} 
\section{Examples}\label{exam}
In order to investigate the effect of the Unruh channel on (EUR) we consider some examples:
\subparagraph{Bell diagonal state} 
To investigate the influence of Unruh effect on (EUR) we consider the bipartite system of Alice and Bob, initially being at inertial frame and sharing a Bell-diagonal state of two Dirac field modes
\begin{equation}\label{bell}
\rho^{AB}=\frac{1}{4}(I \otimes I + \sum_{i=1}^{3} r_{i}\sigma_i \otimes \sigma_i)
\end{equation}
where $\sigma_i$ ($i=1,2,3$) are Pauli matrices. This density matrix is positive if $\vec{r}=(r_1,r_2,r_3)$ belongs to a tetrahedron defined by the set of vertices $(-1,-1,-1)$,$(-1,1,1)$,$(1,-1,1)$ and $(1,1,-1)$. Here we consider the special case in which $r_1=1-2p$, $r_2=r_3=-p$, thus the state in Eq. \ref{bell} becomes
\begin{equation}
\rho^{AB}=p \vert \psi^- \rangle\langle \psi^- \vert + \frac{1-p}{2}(\vert \psi^+ \rangle\langle \psi^+ \vert + \vert \phi^+ \rangle\langle \phi^+ \vert),
\end{equation} 
where $\vert \phi^{\pm} \rangle = \frac{1}{\sqrt{2}}[\vert 00 \rangle \pm \vert 11 \rangle]$ and $\vert \psi^{\pm} \rangle = \frac{1}{\sqrt{2}}[\vert 01 \rangle \pm \vert 10 \rangle]$ are the Bell diagonal states. Bob begins to move with an acceleration $a$. It is associated with the effect of Unruh channel on Bob's particle, thus  the evaluated state can be written as
\begin{equation}
\rho^{AI}=\sum_{i=1}^{2}(I \otimes K_i)\rho^{AB}(I \otimes K_{i}^{\dag}).
\end{equation}
Finally Alice does a measurement on her particle and she tells her measurement choice to Bob. Since Alice is in inertial frame, the measurement
outcomes should be independent of the motion of quantum memory $B$. The new post measurement state becomes
\begin{equation}\label{df}
\rho^{OI}=\sum_{i} (\vert o_i \rangle \langle o_i \vert \otimes \mathbf{I_a})\rho^{AB_a} (\vert o_i \rangle \langle o_i \vert\otimes \mathbf{I_a}),
\end{equation}
where $\lbrace \vert o_i \rangle \rbrace$’s are the eigenstates of the measured observable $\hat{O}$. Here, the conditional von Neumann entropy would be changed, since Eq. \ref{df} gives
new post measurement states. 
In Fig.\ref{Fig1} the (EULB) has plotted in terms of the acceleration of Bob for $p=\frac{1}{2}$. Here we consider the two complementary observable $\sigma_x$ and $\sigma_y$. From Fig. \ref{Fig1}, it is obvious that the (EULB) increases with growing the acceleration of Bob. It is due to that the Unruh effect reduces the quantum information between Alice and Bob, thus the information of Bob about the Alice's measurement reduces. As can be seen from Fig. \ref{Fig1} Holevo (EULB) is tighter than Berta's lower bound.  

\subparagraph{Two-qubit $X$ states}As an another example, we consider a special class in which Alice and Bob, initially being at inertial frame and sharing a two qubit
X states of two Dirac field modes
\begin{equation}
\rho^{AB}=p \vert \psi^{+}\rangle \langle \psi^{+} \vert + (1-p)\vert 11 \rangle \langle 11 \vert \vert,
\end{equation}
where $ \vert\psi^{+} \rangle = \frac{1}{2} ( \vert 01 \rangle + \vert 11 \rangle)$ is a maximally entangled state and $0\leq p \leq 1$. Here we consider the case in which $p=1$, i.e. Alice and bob initially share a maximally entangled state. In ordinary uncertainty game (when Bob in inertial frame $a=0$), If the particle, $A$, and quantum memory, $B$, are maximally
entangled Bob can guess both measurement $\hat{Q}$ and $\hat{R}$ perfectly and (EULB) is equal to zero i.e.
\begin{equation}
S(Q\vert B)+S(R \vert B) \geq 0. 
\end{equation}

In Fig. \ref{Fig2} (EULB) is plotted for two-qubit $X$  states with $p=1$. As can be seen from Fig. \ref{Fig2}, when Alice and Bob being in inertial frame (i.e. $a=0$), as expected the Berta and Holevo lower bound are the same and equal to zero. However by increasing the acceleration (EULB) increases. From Fig. \ref{Fig2} one can conclude that Holevo lower bound is tighter than Berta's lower bound for nonzero acceleration. 
\section{CONCLUSION}\label{con}
In ordinary uncertainty game for quantum-memory assisted (EUR), Bob
sends Alice a quantum state. $A$ initially correlated with another 
quantum memory $B$, Alice does a measurement on her particle $A$ and says her measurement choice to Bob, whose task is then to minimize the uncertainty about the measurement outcomes.

Here, we investigate the quantum-memory assisted (EUR) in relativistic framework.  In this work uncertainty game is as follow : First Alice and Bob agree on two observables, $\hat{Q}$ and $\hat{R}$, Bob prepares a particle constructed from
the free fermionic mode in the quantum state and sends it to Alice. After sending, Bob
begins to move with an acceleration $a$, then Alice does a measurement and announces her choice to Bob, whose task is then to minimize the uncertainty about the measurement outcomes. In other words, here we study the effect of the Unruh channel on ordinary uncertainty game. We show that due to Unruh effect the quantum information stored initially in $B$ would be degraded. This leads to an inevitable increasing of the uncertainty on the outcome of Alice's measurements. We also showed that the Holevo (EUR) is tighter Berta (EUR) in non-inertial frame, as it is in inertial frame \cite{Adabi}.

\end{document}